\documentclass[apj]{emulateapj}

\begin{document}

\title{Generalized investigation of the rotation-activity relation:\\
       Favouring rotation period instead of Rossby number}

\author{A.~Reiners\altaffilmark{1}, 
        M.~Sch\"ussler\altaffilmark{2} 
        and V.M.~Passegger\altaffilmark{1}}
\altaffiltext{1}{Universit\"at G\"ottingen, Institut f\"ur Astrophysik, 
  Friedrich-Hund-Platz 1, 37077 G\"ottingen, Germany}
\altaffiltext{2}{Max-Planck Institut f\"ur Sonnensystemforschung,
  Justus-von-Liebig-Weg 3, 37077 G\"ottingen, Germany}



\begin{abstract}

  Magnetic activity in Sun-like and low-mass stars causes X-ray
  coronal emission, which is stronger for more rapidly rotating
  stars. This relation is often interpreted in terms of the Rossby
  number, i.e., the ratio of rotation period to convective overturn
  time. We reconsider this interpretation on the basis of the observed
  X-ray emission and rotation periods of 821 stars with masses below
  1.4\,M$_{\odot}$.  A generalized analysis of the relation between
  X-ray luminosity normalized by bolometric luminosity, $L_{\rm X}/
  L_{\rm bol}$, and combinations of rotational period, $P$, and
  stellar radius, $R$, shows that the Rossby formulation does not
  provide the solution with minimal scatter. Instead, we find that the
  relation $L_{\rm X}/ L_{\rm bol} \propto P^{-2}R^{-4}$ optimally
  describes the non-saturated fraction of the stars. This relation is
  equivalent to $L_{\rm X} \propto P^{-2}$, indicating that the
  rotation period alone determines the total X-ray emission. Since
  $L_{\rm X}$ is directly related to the magnetic flux at the stellar
  surface, this means that the surface flux is determined solely by
  the star's rotation and is independent of other stellar
  parameters. While a formulation in terms of a Rossby number would be
  consistent with these results if the convective overturn time scales
  exactly as $L_{\rm bol}^{-1/2}$, our generalized approach emphasizes
  the need to test a broader range of mechanisms for dynamo action in
  cool stars.

\end{abstract}


%

\section{Introduction}
\label{sect:Introduction}

\begin{figure*}
  \includegraphics[width=.5\textwidth]{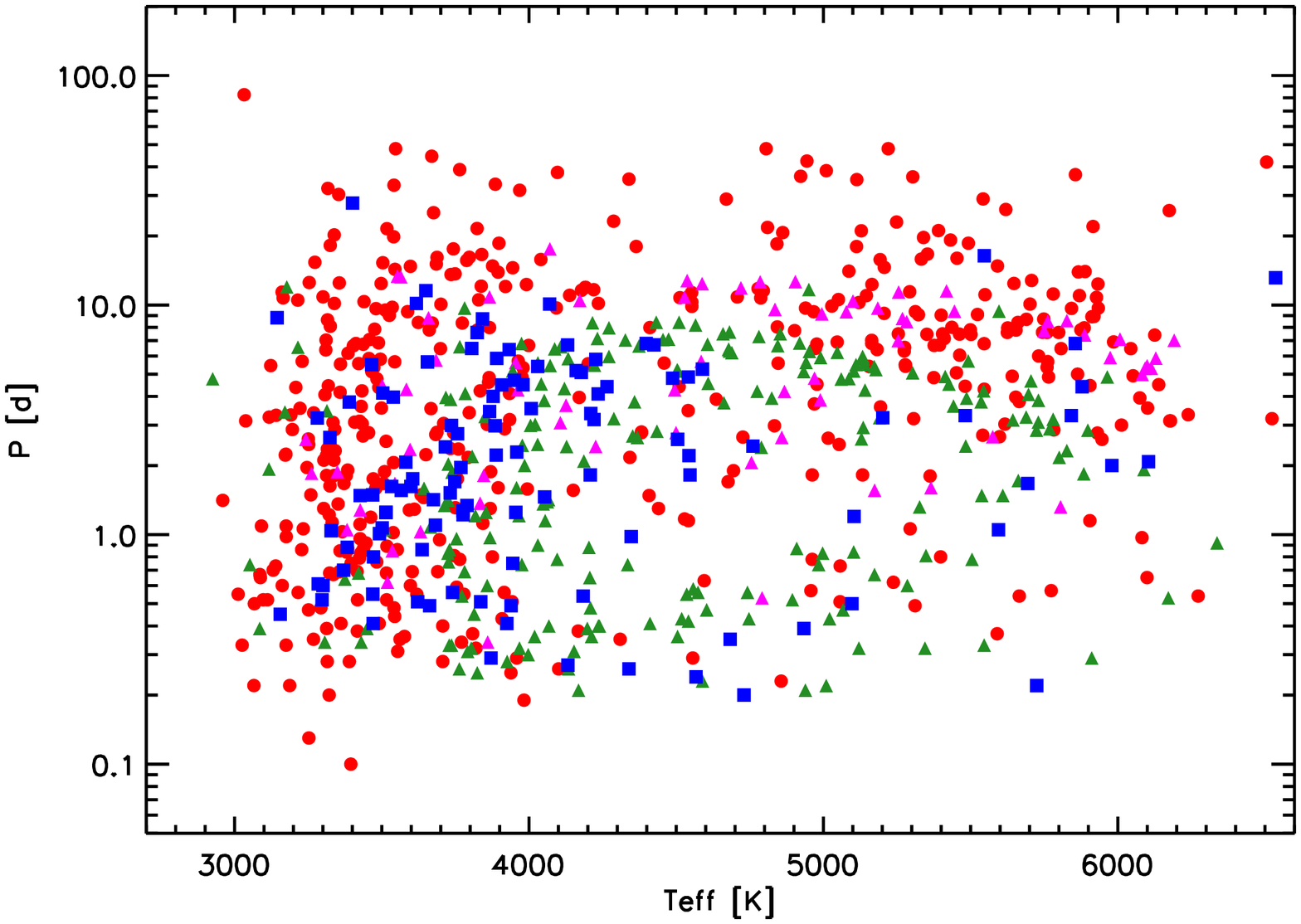}
  \includegraphics[width=.5\textwidth]{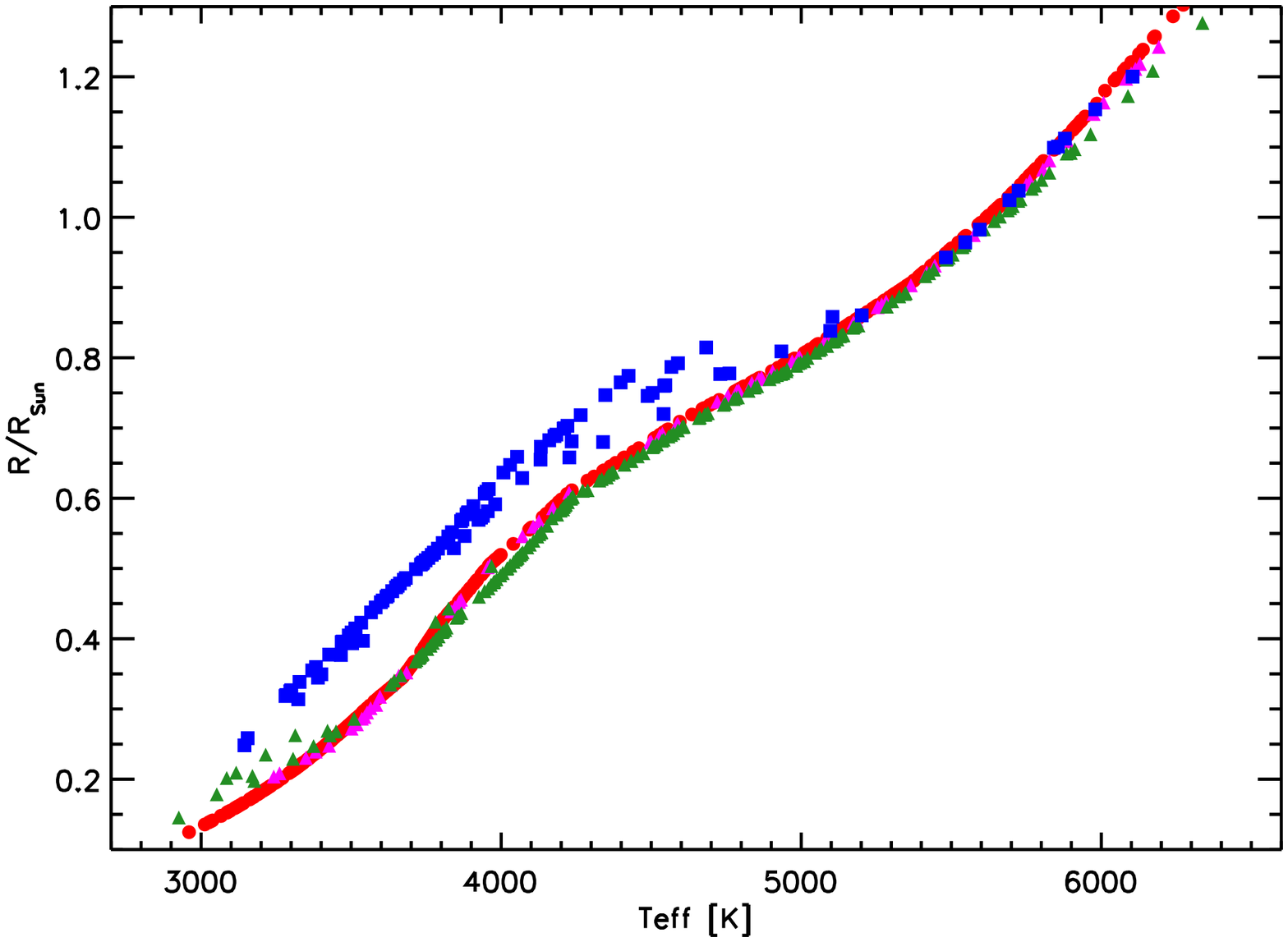}
  \caption{\label{fig:Pteff}Rotation period (left panel) and radius
    (right panel) vs. effective temperature for all sample
    stars. \emph{Blue squares}: very young stars (up to 50\,Myr);
    \emph{green triangles}: young stars (between 85 and 150\,Myr);
    \emph{magenta triangles}: intermediate age stars (600--700\,Myr);
    \emph{red circles}: field stars.}
\end{figure*}

Stars exhibit signs of magnetic activity through prominent emission of
non-thermal radiation from coronal and chromospheric regions
\citep{2003SSRv..108..577F, 2004A&ARv..12...71G, 2008LRSP....5....2H},
and through significant amounts of magnetic flux detectable on their
surfaces \citep{2009ARA&A..47..333D, 2012LRSP....9....1R}. It is
believed that chromospheric and coronal regions are magnetically heated
and that the generation of magnetic energy by large-scale dynamo action
in the stellar convection zone is driven by rotation and convection
\cite[e.g.,][]{2010LRSP....7....3C}.

The relationship between stellar rotation and activity is extensively
discussed in the literature. As soon as the first samples of activity
measurements together with information on stellar rotation became
available, it was realized that magnetic activity is stronger in rapid
rotators, while slowly rotating stars (such as the Sun) exhibit
relatively low levels of non-thermal emission and host only weak average
magnetic fields. \citet{1981ApJ...248..279P} found that the X-ray
luminosity, $L_{\rm X}$, among cool stars (G--M) strongly depends on
rotation rate, with $L_{\rm X} \propto (v_{\rm rot}\,\sin{i})^{1.9 \pm
0.5}$ and no dependence on bolometric luminosity.  The rich dataset from
the Mt~Wilson survey was used by \citet{1984ApJ...279..763N} to study
the relation between chromospheric activity and rotation. They showed
that the ratio of chromospheric flux (observed in Ca\textsc{ii}) to
bolometric flux correlates with the ratio of a color-dependent function
$\tau(B-V)$, which they identified with the convective overturn time, to
the rotation period, $P$. $\tau/P$ is proportional to the inverse Rossby
number, 1/Ro (sometimes called Coriolis number), which describes the
strength of the rotational effect on the convective flows. Under certain
assumptions, the driving of a mean-field $\alpha\Omega$ dynamo is
$\propto {\rm Ro}^{-2}$ \citep{1978GApFD...9..241D,
2001MNRAS.326..877M}, which motivates a scaling of activity in terms of
the Rossby number.

It is clear that comparing $\tau/P$ against the ratio of chromospheric or
coronal flux to bolometric flux (or luminosity) involves the possibility that
bolometric flux and overturn time may actually depend on each other in such a
way that they cancel each other, hence this comparison might only reflect the
relation between non-thermal flux and rotation period alone, i.e.,
$L_{X}/L_{\rm bol}$ vs.\ $\tau/P$ simply reflects the relation $L_{X}$ vs.\
$1/P$. It was pointed out early on, e.g., by \citet{1985ApJ...298..761B} and
\citet{1986LNP...254..184B} that in fact coronal emission and rotation provide
a better correlation, or at least that it cannot be decided which one is the
relevant relation.

The activity-rotation relationship was re-investigated by
\citet{2003A&A...397..147P} on the basis of a large observational sample of
259 field and cluster stars with known X-ray emission and rotation rates. They
showed that two emission regimes exist in cool stars: one for which X-ray
luminosity depends on rotation period, and a second regime where the ratio
between X-ray and bolometric luminosity is constant (called the saturated
regime). They also derived an empirical overturn time, $\tau_{\rm e}$, to
calculate an empirical Rossby number, ${\rm Ro}_{\rm e}=P/\tau_{\rm e}$, for
their sample stars. The function $\tau_{\rm e}(B-V)$ was determined such that
the scatter in the relation between normalized X-ray luminosity, $L_{\rm
  X}/L_{\rm bol}$, and ${\rm Ro}_{\rm e}$ is minimized for the non-saturated
stars. Similar to \citet{1984ApJ...279..763N} they found that $\tau_{\rm e}$
resembles theoretical overturn timescales. However, they also note that
$\tau_{\rm e}$ roughly scales as $L_{\rm bol}^{-1/2}$, which implies that a
relation $L_{\rm X}/L_{\rm bol} \propto {\rm Ro}_{\rm e}^{-2}$ is equivalent
to $L_{\rm X} \propto P^{-2}$, without any dependence on other stellar
parameters.\footnote{Similar results were found for the total Ca\textsc{ii}
  surface flux \citep{1982A&A...107...31M, 1983A&A...121..190C}.}
\citet[][W11]{2011ApJ...743...48W} extended the available data and presented a
sample of 824 cool stars with X-ray luminosities and rotation periods. Similar
to \citet{2003A&A...397..147P}, they determined an empirical overturn time by
minimizing the scatter between normalized X-ray luminosity and a fit in terms
of a power law in the (empirical) Rossby number. For the unsaturated stars in
their sample, they found that $L_{\rm X}/L_{\rm bol}$ depends on Ro with an
exponent of $-2.18 \pm 0.16$ if all stars are used. The authors point out that
the sample suffers from biases due to the selection of only X-ray detected
sources, and they aim to overcome this bias by restricting the fit to an
`unbiased' subsample. From this sample, they found that the slope of the
relation between $L_{\rm X}/L_{\rm bol}$ and Ro has an exponent of $-2.70 \pm
0.13$.

A potential problem of empirically determined convective overturn times
is that the physical meaning of this quantity is rather unclear since
the overturn time varies with depth and it is not clear at which
location the dynamo is most efficient. Furthermore, convective overturn
times are likely to depend on other parameters like, e.g., metallicity,
and change as stars age \citep[e.g.,][]{2010A&A...510A..46L,
2013ApJ...776...87S}.

While it is obvious that one can always derive a function $\tau_{\rm
  e}$ that minimizes the scatter of the activity-rotation relation, it
is not clear whether the similarity of the empirical overturn times
with those resulting from mixing-length models of convection justifies
conclusions concerning the nature of the dynamo process. In this
paper, we therefore reduce the freedom of choosing this function by
assuming that it is a power law of a fundamental stellar parameter
that can, at least in principle, be determined independently for each
sample star. For this we choose the stellar radius, $R$. Similarly, we
assume a power law for the dependence of $L_{\rm X}/L_{\rm bol}$ on
the rotational period, $P$. A fitting procedure based on minimizing
scatter then yields optimal values for the two exponents of the power
laws for the unsaturated part of the stellar sample.  Our main result
is that this procedure leads to a better fit of the data (less
scatter) than the introduction of the empirical Rossby number. The
resulting power-law exponents indicate that $L_{\rm X}\propto 1/P^2$
in the unsaturated regime. Since $L_{\rm X}$ is found to be directly
related to the magnetic flux on the stellar surface
\citep{2003ApJ...598.1387P, 2014arXiv1404.2733V}, this means that the
generation of magnetic flux by the stellar dynamo in cool stars is a
function only of rotation period, independent of other stellar
parameters (such as radius or mass). Above a critical rotation period,
$L_{\rm X}$ saturates at a level of $10^{-3}L_{\rm bol}$ for each
star.  Since our results are free from any assumption about the dynamo
mechanism, they can be used to assess predictions from dynamo theory
in the most general way.

\section{Data}

\subsection{Stellar sample}

We use the comprehensive sample analysed by
W11. It is based on a compilation started by
\citet{2003A&A...397..147P} and was extended by further literature data
and from observations of stellar clusters. In total, the sample
comprises 824 stars, of which we consider the 821 stars with masses
below 1.4\,M$_{\odot}$.

The sample contains stars of very different age, representing the
evolution of stars from a few $10^7$ years up to ages of field
stars. Specifically, the sample contains members of NGC~2547 (60
stars), IC~2602 (28), IC~2391 (13), $\alpha$~Persei (40), the Pleiades
(146), NGC~2516 (14), Praesepe (20), the Hyades (49), as well as 445
field stars.  More details on the references for individual
measurements together with cluster ages and distances are provided in
W11. Here, we adopt ages collected in that
work. In our figures, we distinguish four groups of stellar ages,
which we define as very young (up to 50\,Myr; NGC~2547, IC~2602,
IC~2391), young (between 85 and 150\,Myr; $\alpha$~Persei, Pleiades,
NGC~2516), intermediate (600--700\,Myr; Praesepe and Hyades), and
field stars.

\subsection{Updated stellar parameters}

\subsubsection{Parameters derived from models}

The catalog of W11 contains measured data for rotation, X-ray luminosity, and
color, together with several other indicators. In addition, stellar parameters
such as mass, effective temperature, radius, and depth of the convective
envelope were calculated from the models of \citet{2000A&A...358..593S}. We
updated these parameters by interpolating the tables from
\citet{2000A&A...358..593S}, which led in some cases to values that somewhat
differ from those given in W11. The discrepancy is most pronounced for the
depth of the convective envelope.  This is probably due to the fact that the
parameter range is poorly covered for masses around $M = 0.35$\,M$_{\odot}$ in
the tables of \citet{2000A&A...358..593S}. The minimum mass for the appearance
of a radiative core is $M = 0.35$\,M$_{\odot}$ \citep{1997A&A...327.1039C},
and we set the stars to be fully convective for $M \le 0.35$\,M$_{\odot}$,
thus extending the tabulated points in \citet{2000A&A...358..593S}. For
consistency, we also updated the stellar parameters mass, radius, effective
temperature, and we redetermined bolometric luminosity and used these values
for our analysis.

\subsubsection{Rotation periods}

For some of the stars in the sample, improved measurements of rotation
periods became available during the last years, and some of them
significantly differ from the ones collected in
W11. The new periods used for the analysis in
this paper are listed in Table\,\ref{tab:periods}. 

Figure\,\ref{fig:Pteff} shows an overview of the sample. We plot
rotation period (left) and radius (normalized by the solar radius;
right) versus effective temparature for all sample stars, indicating
ages by different colors and symbols. The sample shows the typical
signatures evident of mass-dependent rotational evolution \citep[see,
e.g.,][]{2010ApJ...721..675B, 2012ApJ...746...43R}.

We note that none of the results shown in the following changes
qualitatively if the original stellar parameters and periods from
W11 are adopted.

\begin{table}
  \begin{centering}
    \caption{\label{tab:periods}Updated rotation periods}
    \begin{tabular}{lrrc}
      \hline
      \hline
      \noalign{\smallskip}
      Star & \multicolumn{2}{c}{Period [d]} & Reference\\
      & Old & Updated & \\
      \noalign{\smallskip}
      \hline
      \noalign{\smallskip}
      GJ 182	& 1.86 & 4.41 & (ks07) \\					
      GJ 494	& 1.54 & 2.889	& (ks07) \\					
      GJ 551	& 42.00 & 82.53	& (ks07) \\
      GJ 2123A	& 7.79  & 0.32	& (ks07) \\					
      GJ 669A	& 19.81 & 0.95	& (ks07) \\					
      HD 95650/GJ410 & 2.94 & 14.81 & (fh00) \\					
      GJ 388/AD Leo  & 2.6 & 2.23 & (en09) \\
      G99-49	& 0.5 & 1.81	& (ir11) \\
      GJ 1156	& 0.87 & 0.491	& (ir11) \\
      GJ 493.1	& 0.21 & 0.6	& (ir11) \\
      GJ 791.2	& 0.32 & 0.346	& (ir11) \\
      \noalign{\smallskip}
      \hline
      \noalign{\smallskip}
    \end{tabular}\\
  \end{centering}
  \emph{References} -- (ks07): \citet{2007AcA....57..149K}; (fh00):
  \citet{2000AJ....120.3265F}; (ir11): \citet{2011ApJ...727...56I};
  (en09): \citet{2009AIPC.1135..221E}
\end{table}

\section{Results}
\label{sect:results}

\subsection{The generalized rotation-activity relation}
\label{subsect:relation}

We assume that the normalized X-ray luminosity, $L_{\rm X}/L_{\rm bol}$,
depends on rotation period and on a combination of parameters given by
the structure of the star, such as mass, radius, temperature, or depth
of the convection zone. Because they all scale in some way with the mass
or the radius of the star, we condense the dependence on stellar
parameters in the radius $R$, and search for an optimal representation
that minimizes the scatter in $L_{\rm X}/L_{\rm bol}$.  For a 
generalized dependence of $L_{\rm X}/L_{\rm bol}$ on $P$ and $R$ in the
form of a combination of power laws,
\begin{equation}
  \label{eq:powerlaw}
  \frac{L_{\rm X}}{L_{\rm bol}} \propto R^{\alpha} P^{\beta},
\end{equation}
we considered fits through the resulting distributions of points in
the plane $\log(L_{\rm X}/L_{\rm bol})$ vs. $\log(kR^{\alpha}
P^{\beta})$, where $k$ represents the constant of proportionality in
Eq.~\ref{eq:powerlaw}. The fit curves are composed of two linear
parts: one for the unsaturated regime, which should follow the
relationship given by Eq.~\ref{eq:powerlaw} and therefore must have a
slope of unity, and another linear fit for the saturated part of the
sample. The best fit is then defined as the one showing the minimum
scatter of the data points with respect to the fit curve.  The
location of the break in the slope of the fit curve between the
unsaturated and the saturated parts is found by including the break
point in the minimization process. 

The linear regression curves are calculated using an adapted version of
a procedure studied by \citet{1990ApJ...364..104I}, who provide
algorithms for five methods for linear regression fits to bivariate
data.\footnote{\texttt{idlastro.gsfc.nasa.gov/ftp/pro/math/sixlin.pro}}
The different methods are useful for taking into account different
distributions of uncertainties in the data. The `standard' way of fitting a
slope in a variable Y to another variable X is the ordinary least
squares method, OLS(Y$|$X). This method assumes that Y measures (with
uncertainty) the value of a parameter that depends on a known variable X
(with no or very little uncertainty). A second method, OLS(X$|$Y),
calculates the fit under the assumption that Y is the variable that is
well defined and the scatter in the observed sample is due to a
distribution (or measurement uncertainties) in X. A third method, the
OLS bisector, performs the fit under the assumption that the scatter in
the distribution is due to scatter in both variables X and Y with
symmetrically distributed scatter. This relation provides the regression
that produces the 'best-looking' fit, i.e., a line that lies closest to
all data points. In our situation, typical uncertainties in the
variables X and Y are not symmetric. The intrinsic scatter in $L_{\rm
X}$ of a sun-like star due to stellar variability is typically on the
order of a factor of 2 but higher during flare events
\citep{1995ApJ...450..392S}. Uncertainties in $L_{\rm bol}$ and
differences between X-ray calibrations from different instruments can
add more uncertainty so that a factor of two is probably a lower limit
for the uncertainty in $L_{\rm X}/L_{\rm bol}$ in our sample. Stellar
radii are derived in a consistent way, so that residual errors are
probably only a few percent. The rotation periods in the sample are also
relatively well constrained. We do not expect any systematic differences
between periods measured by different authors, and the variability of
observed rotation periods in sun-like stars is reported to be on the
order of 10--20\,\% or lower \citep{1996ApJ...466..384D}, i.e., the
uncertainty in our variable X is a factor of 10--20 less than the
uncertainty in our variable Y. Thus, we conclude that OLS(Y$|$X) is the
appropriate method for the calculation in our case. We thereby ignore
the uncertainty in $P$, which probably leads to a slight underestimate
of the slope \citep{1990ApJ...364..104I}. We estimate this effect to be
less than 0.1 from tests using the OLS(Y$|$X) and OLS bisector methods.

For the optimal solution with a scatter of 0.346\,dex we find values
of $\alpha = -4.3$, $\beta = -2.2$, $k =
1.86\times10^{-3}\,$d$^{-\beta}$\,R$_{\odot}^{-\alpha}$, and the break
point at $\log(L_{\rm X}/L_{\rm bol})=\log(kR^{\alpha}P^{\beta}) =
-3.14$. Up to 3 digits, the scatter is the same also for the
combination $\alpha = -4$ and $\beta = -2$. For simplicity, we shall
consider these values as our optimal solution in what
follows.\footnote{For the original sample, i.e., if we use all periods
  and luminosities reported in W11, we find
  $\alpha = -4.1$, $\beta = -2.0$, $k =
  1.23\times10^{-3}\,$d$^{-\beta}$\,R$_{\odot}^{-\alpha}$, and the
  break point at $\log(L_{\rm X}/L_{\rm
    bol})=\log(kR^{\alpha}P^{\beta}) = -3.20$.}

\begin{figure}
  \centering 
    \includegraphics[width=.47\textwidth]{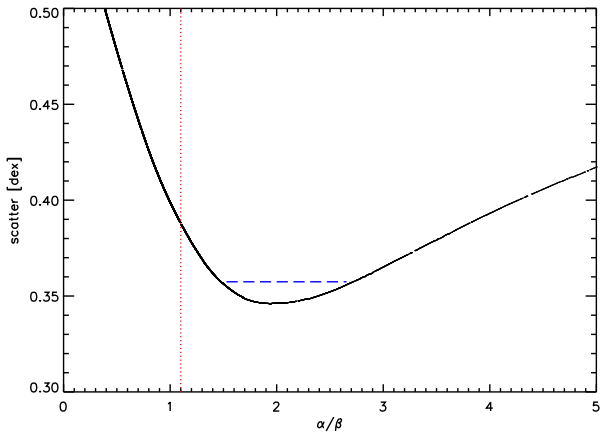}
    \caption{Scatter (1\,$\sigma$) of $L_{\rm X}/L_{\rm bol}$ around
      Eq.\,\ref{eq:powerlaw} for different values of the ratio
      $\alpha/\beta$.  The blue dashed line marks the
      2$\sigma$-interval for $\alpha/\beta$ assuming that the
      statistical uncertainty in the measurement of $L_{\rm X}/L_{\rm
        bol}$ equals the minimum scatter 0.346\,dex. The red dotted
      line shows the approximate location derived in W11 from the
      Rossby-formulation based on their 'unbiased' subsample of stars
      (see text).}
     \label{fig:alphabeta}
\end{figure}

In order to determine a confidence interval for the optimal solution, we note
that the nature of the assumed power-law relationship in the unsaturated part
implies that the scatter of the data points around Eq.\,\ref{eq:powerlaw} on
the log-log plane, and hence the quality of the fit, depends only on the ratio
of the exponents, $\alpha/\beta$.  We thus calculated the scatter as a
function of $\alpha/\beta$ by determining the standard deviation of the data
points from the broken power-law fit, using the value of the break point that
minimizes the standard deviation for each given value of $\alpha/\beta$. We
carried out this procedure for 101 values for each of the parameters $\alpha$
and $\beta$ with $-6\le\alpha\le0$ and $-6\le\beta\le 0$. The resulting
relation between the scatter (in dex) and $\alpha/\beta$ is shown in
Fig.\,\ref{fig:alphabeta}.  We find a smooth distribution with a minimum
located at $\alpha/\beta = 1.93$ (2.06 for the original sample). We estimate a
confidence interval of this solution assuming that the uncertainties in
$L_{\rm X}/L_{\rm bol}$ follow a normal distribution around our fit. In other
words, we assume that the uncertainty of each individual measurement is equal
to the minimum of the scatter. With a minimum scatter of 0.346\,dex and a
total number of 349 measurements of stars in the unsaturated regime, we
estimate \citep[see][]{1986nras.book.....P} as 2$\sigma$ confidence interval
of the ratio $\alpha/\beta$ the range
\begin{equation}
  1.53 < \frac{\alpha}{\beta} < 2.66, 
  \label{eq:alphabeta}
\end{equation}
which is indicated by the blue dashed line in Fig\,\ref{fig:alphabeta}. For
the individual values of $\alpha$ and $\beta$ we find the $2\sigma$ intervals
\begin{eqnarray}
  -5.04 < & \alpha & < -3.54,\\
  -2.37 < & \beta  & < -1.88. 
\end{eqnarray}
We note that the assumption of normally distributed uncertainties in
the individual measurements of $L_{\rm X}/L_{\rm bol}$ is probably not
entirely correct. Nevertheless, a 1\,$\sigma$ uncertainty of
0.346\,dex (more than a factor of 2) appears to be a realistic
estimate that captures several systematic effects, such as instrument
differences, flares (occuring statistically), and contamination from
binaries. Ratios outside the confidence interval can therefore be
regarded as statistically unlikely, even if we do not fully understand
the sources of the measurement uncertainties. The dotted red vertical
line in Fig.\,\ref{fig:alphabeta} indicates the value of
$\alpha/\beta=1.1$ that we infer for the formulation of
W11 in terms of the Rossby number (see
Sect.~\ref{subsect:Rossby} below); it is clearly outside the 2$\sigma$
confidence interval.

Our result provides information on the dependence between the parameters used
in Eq.\,\ref{eq:powerlaw}. It is important to realize that in the available
sample, rotation, color, mass, radius, age, etc.\ are severely degenerate, for
example because very few sun-like stars are rapidly rotating. Our result is
that rotation period, $P$, and radius, $R$ can explain the existing activity
observations but underlying physical relations may be hidden by sample
degeneracies. Further data on stars occupying sparsely populated parameter
ranges would help to remove this degeneracy.

\subsection{Comparison to \citet{2011ApJ...743...48W}}
\label{subsect:Rossby}

\begin{figure*}[ht!]
  \centering \mbox{
    \includegraphics[width=.47\textwidth]{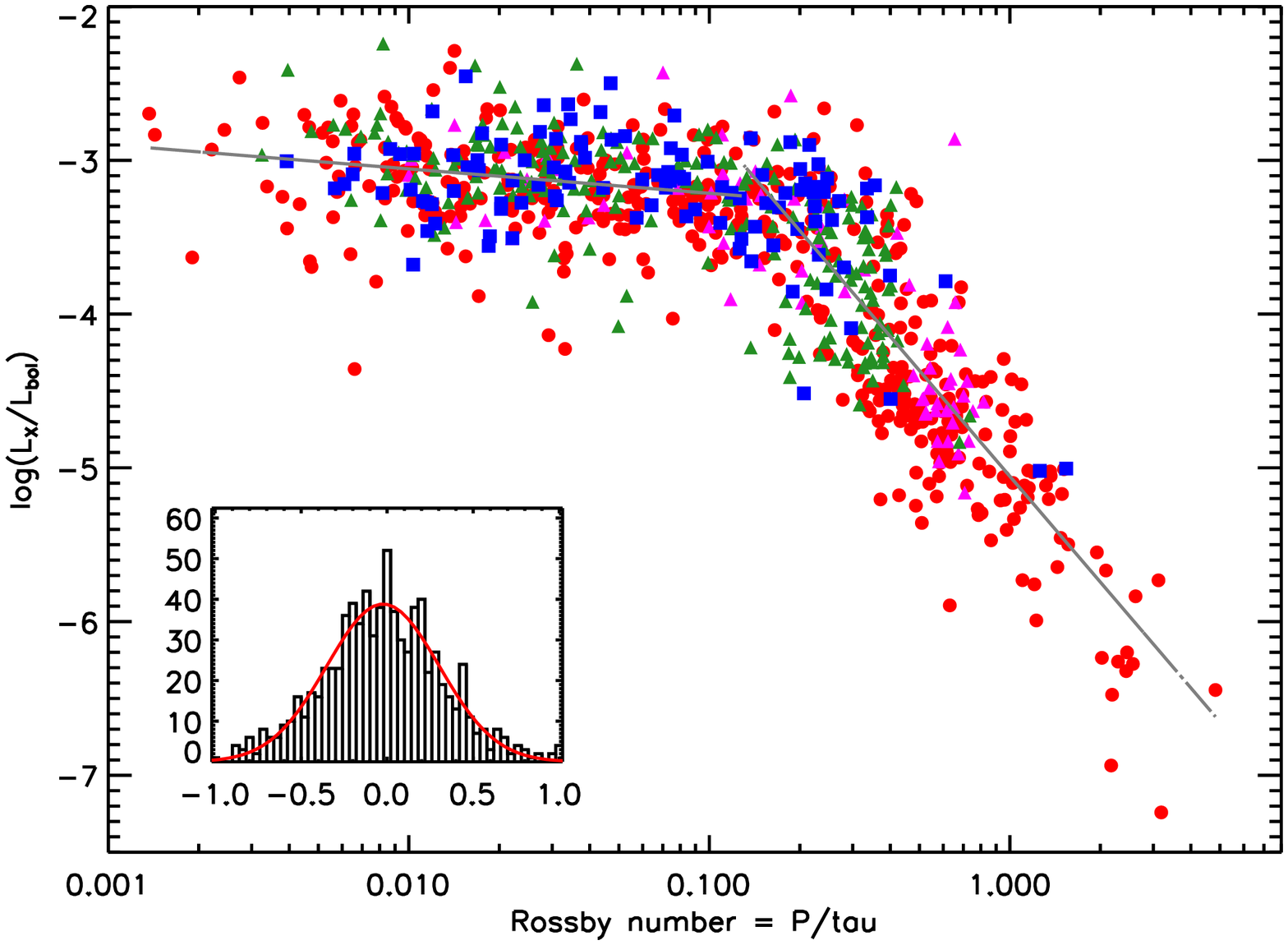}
    \includegraphics[width=.47\textwidth]{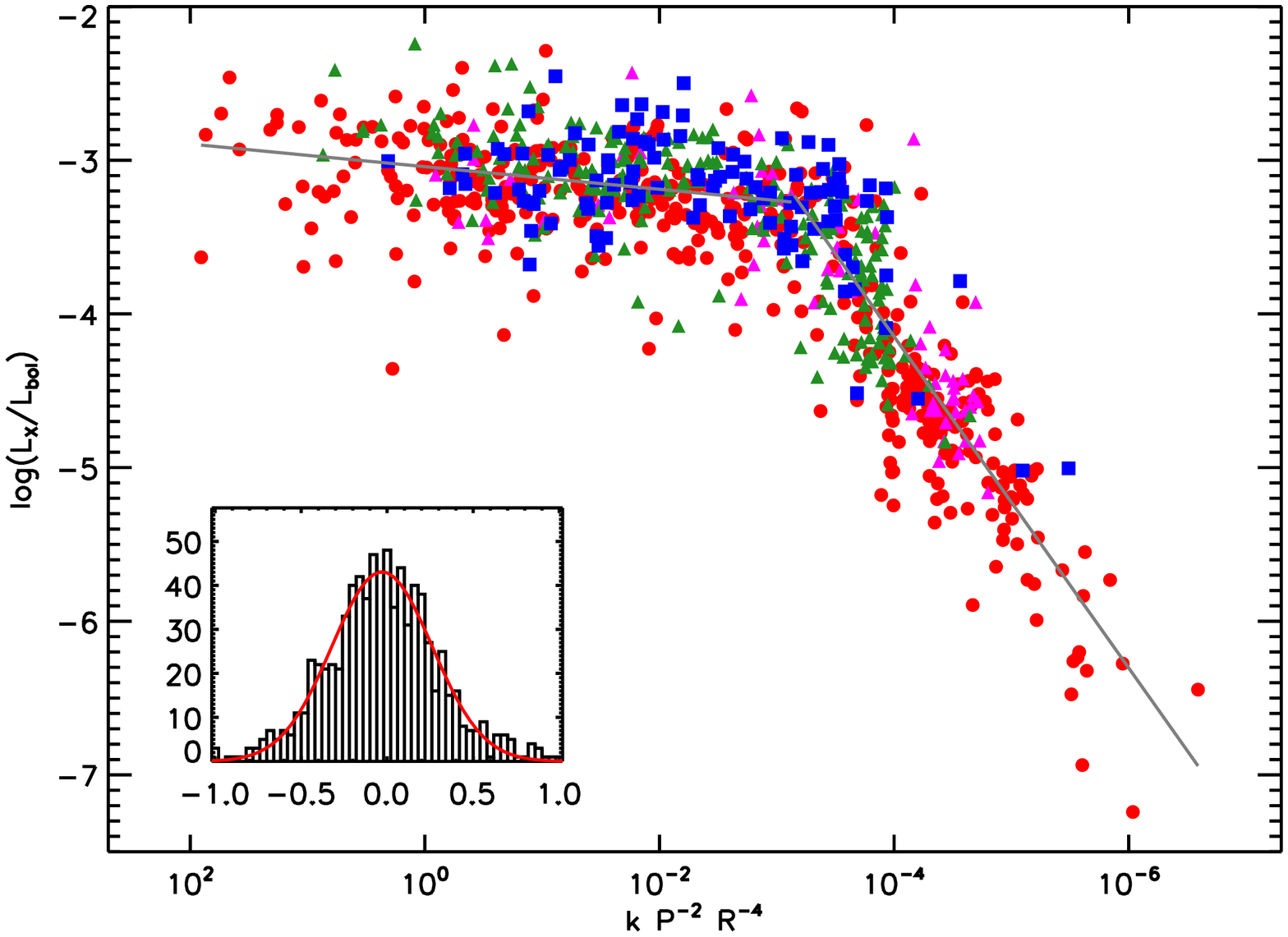}}
  \caption{\label{fig:lxlbol}Normalized X-ray luminosity as a function
    of empirical Rossby number (left) and as a function of $k P^2 R^4$
    (right). Different colored symbols show stars at different ages
    (see Fig.\,\ref{fig:Pteff}). The distributions are fit using a
    broken power-law (grey lines) with the break at ${\rm Ro} = 0.13$
    (left panel) and $\log{kP^{-2}R^{-4}} = -3.14$ (right
    panel). Scatter around the fit is shown in the inset of each
    panel, a Gauss fit is overplotted in each panel's inset as a red
    line.}
\end{figure*}

\begin{table}
  \begin{centering}
    \caption{\label{tab:Wright_compare}Slope $\gamma$ in the unsaturated regime
    for the three cases in W11}
    \begin{tabular}{cccc}
      \hline
      \hline
      \noalign{\smallskip}
      Sample & OLS(Y$|$X) & OLS bisector & OLS(X$|$Y) \\
      \noalign{\smallskip}
      \hline
      \noalign{\smallskip}
      $0.13 <$ Ro & $-2.16 \pm 0.08$ & $-2.58 \pm 0.08$ & $-3.17 \pm 0.12$\\
      $0.2 <$ Ro $< 3.0$  & $-2.36 \pm 0.09$ & $-2.90 \pm 0.09$ & $-3.71 \pm 0.17$\\
      `unbiased' sample$^{*}$  & $-1.91 \pm 0.18$ & $-2.24 \pm 0.16$ & $-2.67 \pm 0.25$\\
      \noalign{\smallskip}
      \hline
      \noalign{\smallskip}
    \end{tabular}\\
    $^{*}$ valid in the range $0.3 <$ Ro $< 3$; HD\,81809 is not in the catalog
  \end{centering}
\end{table}

Wright et al.\ used the same sample of stars to obtain a power-law
relation between $L_{\rm X}/L_{\rm bol}$ and an empirical Rossby
number by considering the convective overturn time, $\tau$, as an
adjustable function of stellar mass. By minimizing the scatter of the
data points with respect to the power-law fit they determined the
slope $\gamma$ in
\begin{equation}
  \frac{L_{\rm X}}{L_{\rm bol}} \propto \left(\frac{P}{\tau}\right)^{\gamma}
  \label{eq:Wrightfit}
\end{equation}
in the unsaturated regime. Including all 824 stars in their sample,
they report $\gamma = -2.18 \pm 0.16$ for the best fitting slope. The
authors argue that an OLS bisector fit is appropriate for this
sample. We question this choice because of the asymmetric distribution
of the uncertainties (see Section\,\ref{sect:results}). To compare the
results from different fit methods, we re-calculated the slope using
three different methods, OLS(Y$|$X), OLS bisector, and OLS(X$|$Y). The
results are shown in Table\,\ref{tab:Wright_compare}.\footnote{For
  this comparison, we use the original values for $P$ and $L_{X}$ as
  reported in W11.}  We include three subsamples that are mentioned in
W11, the full sample (with Ro\,$>0.13$), a sample with Ro\,$> 0.2$,
and their 'unbiased' sample (see below). We observe rather large
inconsistencies between our results and their findings, which may be
partly because we cannot reconstruct the overturn times W11 used for
their fit. These overturn times follow the relation from
\citet{2003A&A...397..147P} but W11 convert $V-K$ into $B-V$ based on
a relation that is not provided for all stars. We therefore used the
convective overturn times as determined in W11 as a function of mass,
which do not significantly differ from those of Pizzolato et al.\ as
shown in W11.

An important point in the results from W11 is their construction of an
`unbiased' sample. The authors argue that X-ray bright sources are
easier to detect and may therefore be overrepresented with respect to
stars that are X-ray dim. This could be particularly important for the
mean $L_{X}$ for a given value of $P$ because slow rotators with low
X-ray brightness may be systematically underrepresented. To construct an
X-ray complete sample, they chose the 36 stars from
\citet{1996ApJ...466..384D} for which rotation periods could be measured
over five or more seasons. According to W11, the 36 stars all show
measurable X-ray emission and W11 argue that the sample therefore does
not suffer from X-ray luminosity bias.\footnote{We note that HD\,81809
is a member of the 36 stars from \cite{1996ApJ...466..384D} but we could
find it neither in the sample of W11 nor in the NEXXUS database
\citep{2005ESASP.560..755L}.} We do not agree that this particular
sample should be less biased than others, in particular because it is a
sample of 36 selected out of 100 stars based on detectability of
photometric periods. We calculated the slope for this sample and include
our results in Table\,\ref{tab:Wright_compare}. We cannot reproduce the
value of $\gamma = -2.7$ for the OLS bisector method. We revisit the
question of luminosity bias in Section\,\ref{sect:bias}.

For completeness, we compare the scaling reported in W11 ($\gamma =
-2.7$) to our solution. We note that this is not comparing the same
samples, but we find it helpful to discuss the $\gamma = -2.7$ scaling
to our generalized results. The convective overturn time for the W11
solution was parametrized in terms of stellar mass, $M$, by a
second-order double-logarithmic polynomial (see their Eq.\,11). For
the stars considered in the unsaturated part of the sample, the values
of $M/M_\odot$ and $R/R_\odot$ are almost identical, so that we can
replace $M/M_\odot$ in their relation by $R/R_\odot$ without
introducing significant scatter. Furthermore, we can approximate their
Eq.\,(11) by the first-order relation
\begin{equation}
  \log{\tau} = 1.19 - 1.08 \log{M/M_\odot},
  \label{eq:tau_Wright}
\end{equation}
so that we can rewrite Eq.\,(\ref{eq:Wrightfit}) 
in the form
\begin{equation}
  \frac{L_{\rm X}}{L_{\rm bol}} \propto
  \left(\frac{R^{-1.08}}{P}\right)^{2.7} = P^{-2.7} R^{-2.9},
  \label{eq:Wrightapprox}
\end{equation}
which is equivalent to Eq.\,(\ref{eq:powerlaw}) with
$\alpha/\beta=2.9/2.7=1.1$. This value is indicated by red dotted line
in Fig.\,\ref{fig:alphabeta}; it is outside the 2$\sigma$ confidence
interval for the optimal value of $\alpha/\beta$. The expected scatter
for $\alpha / \beta = 1.1$ is approximately 0.38\,dex. A more direct
comparison of the fit qualities can be achieved if we use the original
Rossby scaling provided by W11 and compare its scatter to the one
determined from the distribution using $\alpha = -4$ and $\beta =
-2$. In Fig.\,\ref{fig:lxlbol}, we show normalized X-ray luminosity as
a function of Rossby number (left panel) and as a function of $k
P^{-2} R^{-4}$ for minimal scatter (right panel). In both panels, we
overplot broken power-law fits to the data. For the values from W11,
we break the power law at ${\rm Ro}_{\rm sat} = 0.13$ as reported in
that work. For the values of $k$ and the power law break in the
$\alpha = -4$ and $\beta = -2$ solution we take the values of the
optimal solution.

From both descriptions, we determined the scatter around the power-law
fit by calculating the standard deviations of the distributions shown in
the insets of the panels of Fig.\,\ref{fig:lxlbol}. For the Rossby
formulation, the standard deviation is $\sigma = 0.371$\,dex while for
the generalized formulation (right panel in Fig.\,\ref{fig:lxlbol}) we
have $\sigma = 0.346$\,dex (as found in
Sect.\ref{subsect:relation}). Compared to the simplified approach using
Eq.~(\ref{eq:tau_Wright}), the value for the optimized Rossby
formulation is closer to, but still considerably higher than, the
minimum from our generalized formulation.

\subsection{Comparison to \citet{2003A&A...397..147P}}
\label{subsect:Pizzolato}

A quadratic dependence of X-ray luminosity on rotation rate alone was
already suggested by \citet{1981ApJ...248..279P}. More recently,
\citet{2003A&A...397..147P} pointed out that the empirical turnover
time approximately scales as $\tau\propto L_{\rm bol}^{-1/2}$. With a
rotation-activity relation of the form $L_{\rm X}/L_{\rm bol}\propto
{\rm Ro}^{-2}$ they then obtain
\begin{equation}
 \frac{L_{\rm X}}{L_{\rm bol}} \propto {\rm Ro}^{-2} 
 \propto \frac{1}{P^2 L_{\rm bol}}, 
 \label{eq:pizzolato}
\end{equation}
which is equivalent to $L_{\rm X} \propto P^{-2}$ .  This relation is
consistent with the result of our generalized approach, which can be
seen as follows. Since $L_{\rm bol} \propto R^2 T_{\rm eff}^4$, and also
for the non-saturated stars approximately $T_{\rm eff}\propto R^{1/2}$,
we find with $\alpha=-4$ and $\beta=-2$ the relation
\begin{equation}
  \label{eq:r2t4lbol}
  \frac{L_{\rm X}}{L_{\rm bol}} \propto P^{-2} R^{-4} 
  \propto \frac{1}{P^{2} (R^{2}T^{4})} \propto \frac{1}{P^{2} L_{\rm bol}},
\end{equation}
which is identical to the relation in Eq.~(\ref{eq:pizzolato}).  The
factor of $R^{-4}$ effectively compensates the normalization by the
bolometric luminosity, indicating that the normalization is in fact
unwarranted in the unsaturated regime.

Since we have the values of $L_{\rm bol}$ for the stars in our sample, we can
also directly consider the relation $L_{\rm X}/L_{\rm bol} \propto 1/(P^2
L_{\rm bol})$. It turns out that the scatter from this relation is 0.340\,dex,
which is even slightly lower than the minimum scatter of 0.346\,dex for the
relation given in Eq.~\ref{eq:powerlaw}. This supports the conclusion that the
relation between $L_{\rm X}$ and $P^{-2}$ is better constrained than the
relation between $L_{\rm X}/L_{\rm bol}$ and some combination of $P$ and other
stellar parameters.

\subsection{Saturation}
\label{subsect:saturation}

\begin{figure}
  \centering 
    \includegraphics[width=.47\textwidth]{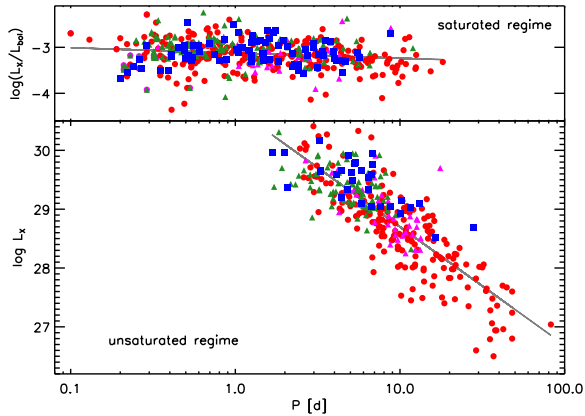}
    \caption{\label{fig:LX_P}Separate representations of X-ray
      activity vs. rotation period in the unsaturated (lower part) and
      saturated (upper part) regimes. Activity is represented in terms
      of $L_{\rm X}$ in the unsaturated regime and in terms of $L_{\rm
        X}/L_{\rm bol}$ in the saturated regime.}
\end{figure}

We have seen that the total X-ray luminosity in the unsaturated regime
scales with $P^{-2}$ and does not depend on any other stellar
parameter. When the total X-ray luminosity reaches a critical value of
about $10^{-3} L_{\rm bol}$, it does not grow further for shorter
rotational periods, i.e., activity saturates. The critical period at
which saturation sets in therefore depends on the stellar luminosity,
$L_{\rm bol}$. We can use our optimal solution to determine the value
of the critical period. For $\alpha = -4$, $\beta = -2$, and
$k=1.86\times10^{-3}$\,d$^{2}$\,R$_{\odot}^{4}$, saturation sets in at
about $\log{k P^{-2} R^{-4}} = -3.14$. With $L_{\rm bol}\propto R^4$
and $L_{\odot}=3.853\cdot10^{33}\,$erg$\,$s$^{-1}$ we find
\begin{equation}
  \label{eq:Psat}
  P_{\rm sat} \, {\rm [d]}= 
    1.6 \left(\frac{L_{\rm bol}}{L_{\odot}}\right)^{-1/2} = 
         \left(\frac{L_{\rm bol}}{1.1\cdot10^{34}}\right)^{-1/2}
\end{equation}
where $L_{\rm bol}$ is in units of erg\,s$^{-1}$. This result is
similar to Eq.\,(6) of \citet{2003A&A...397..147P}. 

For this value of the critical period, we show the distribution of
$\log L_{\rm X}$ vs.\ $P$ for the unsaturated regime together with
$\log(L_{\rm X}/L_{\rm bol})$ vs.\ $P$ for the saturated regime in
the lower and upper panels of Fig.\,\ref{fig:LX_P}, respectively. For
the unsaturated regime, we find the relation
\begin{equation}
  \label{eq:LxPslope}
  \log{L_{\rm X}} = (30.71 \pm 0.05) - (2.01 \pm 0.05) \log{P},
\end{equation}
which is consistent with our optimal value $\beta = -2$.\footnote{For the
  original sample, we find $\beta = 1.97 \pm 0.08$.}

\subsection{A slope in the saturated regime}
\label{subsect:slope}

All three representations shown in Figs.~\ref{fig:lxlbol} and
\ref{fig:LX_P} indicate a slight slope of the rotation-activity
relationship in the saturated regime, i.e. some remnant dependence of
the activity on rotation (or other parameters) even for very rapidly
rotating stars.  Quantitatively, we find for the different representations:
\begin{eqnarray*}
  \log{\frac{L_{\rm X}}{L_{\rm bol}}} &=& (-3.37 \pm 0.06) - (0.16 \pm 0.03) \log{\rm Ro},\\
  \log{\frac{L_{\rm X}}{L_{\rm bol}}} &=& (-3.04 \pm 0.02) - (0.07 \pm 0.01) \log{(kP^2R^4)},\\
  \log{\frac{L_{\rm X}}{L_{\rm bol}}} &=& (-3.12 \pm 0.01) - (0.11 \pm 0.03) \log{P}.
\end{eqnarray*}
There is a statistically significant slope in all three cases. The slope is
least significant (but still above $3\sigma$) in the parameterization with
$P$, while at $\geq5\sigma$ and more in the two other cases. The slope is
likely due to a remaining dependence of the dynamo on rotation period even
when saturation is reached, but it may also be influenced by small differences
in the saturation level between stars of different mass.

\subsection{Luminosity bias}
\label{sect:bias}

\begin{figure}
  \centering 
    \includegraphics[width=.47\textwidth]{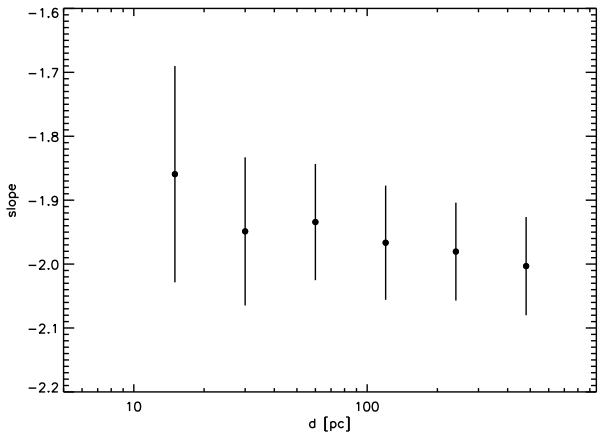}
    \caption{\label{fig:lumbias}Slope $\beta$ for subsamples that contain
      stars out to a maximum distance. Error bars show 1$\sigma$
      uncertainties.}
\end{figure}

W11 pointed out that the slope of the rotation-activity relation may
suffer from a luminosity bias. A possible consequence is that the least
X-ray bright stars are systematically missed, so that the average X-ray
luminosity among the least active stars (the slowest rotators) is
overestimated. This would lead to a slope shallower than the true
relation. We are lacking a statistically unbiased, complete sample of
stars with X-ray and rotation period measurements. Nevertheless, the
large sample of targets allows us to test whether the slope $\beta$ that
we derive in the unsaturated regime depends on the distance to the stars
in the sample. A luminosity bias would be less pronounced in a sample of
nearby stars and become more important if we include increasingly
distant objects. We carried out this test computing the slope $\beta$
for stars in the sample with distances out to 15, 30, 60, 120, 240, and
480\,pc. The results are shown in Fig.\,\ref{fig:lumbias}. We find no
significant trend of $\beta$ as a function of distance limit. There is a
marginal trend towards \emph{higher} absolute values of the slope at
large distances, but it is dominated by the sample limited to 15\,pc,
which has large uncertainties. We conclude that our results do not show
evidence of a luminosity bias.

\section{Discussion}
\label{sect:discussion}

The result of our generalized analysis of the rotation-activity
relation can be summarized as follows: The total X-ray luminosity
scales with rotation period ($P^{-2}$) as long as stellar activity is
not saturated, and X-ray activity saturates for a given star when
$L_{\rm X}/L_{\rm bol}$ reaches a level of about $10^{-3}$. In the
unsaturated regime, this description is equivalent to a scaling
$L_{\rm X}/L_{\rm bol}\propto P^{-2}R^{-4}$, which could be written as
a Rossby number scaling of the form $L_{\rm X}/L_{\rm bol} \propto
{\rm Ro}^{-2}$ if the convective overturn time scales as $\tau\propto
R^{-2}\propto L_{\rm bol}^{-1/2}$. Furthermore, for a given star in
the saturated regime, $L_{\rm X}/L_{\rm bol}$ still
shows a weak but significant dependence on rotation. In what follows
we discuss some physical implications of these results.

\subsection{$L_{\rm X}\propto P^{-2}$}
\label{subsec:diss_lxpropto}

This relationship means that two stars with the same rotation period
emit the same X-ray luminosity, irrespective of their mass or
radius. Since observations indicate that $L_{\rm X}$ is proportional to
$\Phi_{\rm s}$, the total magnetic flux at the stellar surface
\citep{2003ApJ...598.1387P}, it implies that $\Phi_{\rm s}$ depends only
on the rotation rate, but not on any other stellar parameter.  This is
consistent with a recent study by \citet{2014arXiv1404.2733V} on the
relationship between the large-scale surface magnetic flux determined by
Zeeman Doppler Imaging, $\Phi_{\rm V}$, and X-ray luminosity: the two
relations, $L_{\rm X}$ vs. $\Phi_{\rm V}$ and $L_{\rm X}/L_{\rm bol}$
vs. $\Phi_{\rm V}$, yield nearly the same power-law exponents (1.80 and
1.82, respectively). This means that $\Phi_{\rm V}$ is uncorrelated with
$L_{\rm bol}$ and, therefore, uncorrelated with stellar radius or
mass. Similarly, the scalings with rotation period of $\Phi_{\rm V}$ and
magnetic flux density ($\Phi_{\rm V}$ divided by surface area) show the
same power-law exponent, which means that the rotation rate is
uncorrelated with radius or mass (consistent with our value $\alpha =
-4$).

The absence of a dependence of $L_{\rm X}$ on stellar parameters apart from
rotation means that a bigger star shows the same magnetic surface flux as a
smaller star at the same rotation rate. One would have naively expected that
the bigger volume of the convection zone available for dynamo action would
also lead to more magnetic flux being produced by the dynamo, so that also
more flux emerges at the surface.  Also, bigger scales could imply less
dissipation of the large-scale field by turbulent diffusion, thus effectively
increasing the net driving of the dynamo. On the other hand, the turbulent
magnetic diffusivity probably decreases towards cooler (smaller) stars since
the convective motions are slower; the net effect on dynamo driving is
unclear.

\subsection{Saturation}
\label{subsec:diss_saturation}

The existence of a limiting value $L_{\rm X}/L_{\rm bol}\approx 10^{-3}$
for very fast rotators lends itself to at least three possible
interpretations. 

Firstly, it can be seen as indicating that there is a maximum fraction of the
total energy flux of the star that can be converted into X-ray flux. This
could be related to an upper limit of the efficiency of converting the energy
flux into magnetic energy \citep[e.g.,][]{2009ApJ...692..538R,
  2009Natur.457..167C}, although the relation between X-ray flux and magnetic
energy generation in the convection zone is unclear.

A second possible interpretation can be inferred from the scaling of
the critical rotation period for saturation given in
Eq.~(\ref{eq:Psat}): $P_{\rm sat}\propto L_{\rm bol}^{-1/2}\propto
R^{-2}$. This relation implies that the critical rotation rate,
$2\pi/P_{\rm sat}$, is proportional to the surface area of the star,
so that it can be interpreted as the rotation rate at which the
magnetic surface flux in bipolar regions fills the complete stellar
surface \citep[e.g.,][]{1984A&A...133..117V}. If the X-ray flux
ultimately results from the interaction of surface magnetic flux with
near-surface flows, saturation could be a result of this situation. In
this picture, the filling of the surface would also need to imply that
the total surface magnetic flux is saturated, for which evidence is
reported in \citet{2009ApJ...692..538R}. For a solar-type star,
saturation occurs at $L_{\rm X} \approx 4\cdot10^{30}\,$erg\,s$^{-1}$,
roughly a factor of $10^3$ above the value at activity maxima of the
Sun today, during which the area fraction of active regions is a few
percent. For $L_{\rm X}\propto \Phi_{\rm_s}$
\citep{2003ApJ...598.1387P}, the saturated X-ray luminosity would not
be reached for a Sun fully covered by active regions
\citep[e.g.,][]{1978ARA&A..16..393V, 2000ApJ...545.1074D}; however, it
would be consistent for a steeper relationship, such as $L_{\rm
  X}\propto \Phi_{\rm_s}^{1.8}$ proposed by
\citet{2014arXiv1404.2733V}.

Another interpretation can be given in terms of the nonlinearities that
determine the amplitude of the dynamo-generated magnetic field
(R.H.~Cameron, private communication). The present Sun is located in the
lower part of the unsaturated regime and its differential rotation is
almost invariant during the activity cycle. This means that the magnetic
energy is small compared to the kinetic energy in the differential
rotation, which therefore does not experience a strong back reaction of
the toroidal magnetic field it generates. The field amplitude in the
unsaturated regime is thus determined by a nonlinearity affecting the
generation of the poloidal field (`$\alpha$ quenching'). As the rotation
rate grows along the unsaturated branch, at some stage the magnetic
energy becomes comparable to the kinetic energy of differential
rotation, so that a strong back reaction occurs (`$\Omega$
quenching'). If the corresponding nonlinearity is sufficiently hard, it
results in a quasi-saturated regime that is (almost) independent of the
rotation rate.

\subsection{Dynamo models}
\label{subsec:diss_dynamo}

Given the present state of our understanding of solar and stellar
dynamos \citep[see, e.g.,][]{2010LRSP....7....3C}, drawing a
quantitative connection between the dynamo mechanism and the observed
activity indices is by no means straightforward. It seems
uncontroversial that the toroidal flux is generated (from poloidal flux)
by differential rotation (the $\Omega$-effect) while the poloidal flux
is regenerated by some kind of $\alpha$-effect. The latter could be due
to the action of cyclonic convection on the toroidal field, i.e., the
classical Parker loop, which would bring the Rossby number into
consideration. However, it could also result from the Babcock-Leighton
mechanism, i.e., the systematic tilt of bipolar regions with respect to
the direction of rotation.  This tilt can result from the action of the
Coriolis force on flows along rising flux tubes \citep{Fan:2009}. These
flows are not of a convective nature and thus independent from a Rossby
number.

The combination of $\alpha$-effect and $\Omega$-effect provides the
driving of the dynamo, but the relation of this driving to rotation is
uncertain. Since the $\alpha$-effect is due to the action of the
Coriolis force, one can argue that it should be proportional to the
rotation rate, at least for not too rapid rotation. The relation of
differential rotation to the overall rotation rate is much more unclear,
so that the scaling of dynamo driving (expressed by a non-dimensional
dynamo number involving the product of $\alpha$- and $\Omega$-effect)
with rotation is rather uncertain. The dependence on other stellar
parameters is unclear as well and mostly addressed by crude dimensional
arguments.

The scaling of the amplitude of the dynamo-generated field with the
driving dynamo number depends crucially on the kind of nonlinear
backreaction of the magnetic field on its sources, which limits the
growth of the magnetic energy. There are various nonlinearities that
could play a role here, e.g., quenching of $\alpha$-effect and
differential rotation, flux loss by magnetic buoyancy, driving of
large-scale flows, all of which are not well understood quantitatively.
As a consequence, dynamo models mostly consider them in an ad-hoc
fashion.  Keeping in mind these severe uncertainties, it is interesting
to note that at least some models show roughly similar dependencies of
the field amplitude on the dynamo driving (dynamo number, $N_{\rm D}$):
$\propto N_{\rm D}^{1/2}$ for simple one-dimensional models
\citep{1972A&A....20....9S, 1989A&A...223..343S}, and $\propto N_{\rm
D}^{0.65}$ for a state-of-the-art 2D flux-transport dynamo model
\citep{2014arXiv1402.1874K}; the latter is also consistent with the
result from an earlier model by \citet{2010A&A...509A..32J}. However,
quite different dependencies are found with other models
\citep[e.g.,][]{1982A&A...108..322R} and a dynamo model with
$\alpha$-effect and differential rotation taken from global 3D
simulations even fails to show an increase of activity with rotation
rate \citep{2013ApJ...775...69D}.

Eventually, the observed activity indices (such as $L_{\rm X}$) are
related to the surface flux emerging in bipolar magnetic regions. How
this flux is connected quantitatively with the general dynamo amplitude,
i.e., the amount of magnetic flux or magnetic energy generated in the
convection zone, depends on the detailed processes leading to flux
emergence. Again, these processes are not well understood and different
mechanisms are possible: instability of strong magnetic flux tubes
\citep{Fan:2009} or buoyant local flux concentrations compressed to
super-equipartion by turbulent convective flows
\citep{2011ApJ...739L..38N,2014SoPh..289..441N}.

\section{Concluding remarks}
\label{sect:conclusion}

We used activity and rotation measurements of 821 stars, compiled in the
sample of W11, to perform a generalized analysis of the rotation-activity
relation. The sample was updated with recent period measurements from the
literature. We considered the relation of normalized X-ray luminosity,
${L_{\rm X}/L_{\rm bol}}$, on rotation period, $P$, and stellar parameters
condensed in the radius, $R$, in the functional form $R^\alpha P^\beta$. In
the unsaturated regime, we found the representation with the least scatter for
$\alpha=-4$ and $\beta=-2$. Since approximately $L_{\rm bol}\propto R^4$, this
solution is equivalent to $L_{\rm X} \propto P^{-2}$, a relation that had
already been found by \citet{1981ApJ...248..279P} and
\citet{2003A&A...397..147P}. On the other hand, the most recent
parametrization of the activity-rotation relation in terms of the Rossby
number by W11 does show a significantly higher scatter.

In the subsample of stars that are in the saturated regime, we found a
slight but significant growth of ${L_{\rm X}/L_{\rm bol}}$ with faster
rotation. It is unclear whether this trend is due to an effect of $P$
beyond the saturation limit or results from a degeneracy between stellar
mass and rotation rate reflecting that $L_{\rm X}/L_{\rm bol}$ is
slightly larger for more luminous stars.

Formally, we can rewrite $R^{-4}P^{-2}$ in terms of a Rossby number,
${\rm Ro}=P/\tau$, if the convective overturn time scales as
$\tau\propto R^{-2} \propto L_{\rm bol}^{-1/2}$. We then obtain ${L_{\rm
X}/L_{\rm bol}}\propto {\rm Ro}^{-2}$. Unless we have a reliable
determination of the Rossby number in stellar convection zones (provided
that this is possible at all), we cannot decide whether the physical
mechanism behind the rotation-activity relation is better represented by
a Rossby-number scaling of ${L_{\rm X}/L_{\rm bol}}$ or by a purely
rotational scaling of $L_{\rm X}$ independent of other stellar
parameters. However, since the latter involves no assumptions on
physical conditions to be valid in the star, we favour this description
and find it unnecessary to explain the rotation-activity relation in
terms of the Rossby formulation. 

The dependence of dynamo driving (as possibly expressed by a dynamo number),
its nonlinearity, and the amount of magnetic flux emerging at the stellar
surface on rotation and other stellar parameters (such as radius) is quite
unclear. The various factors may well combine to a result that makes the
surface flux independent of stellar radius as indicated by our analysis. One
possible form to condense this independence into a physical model is to assume
a Rossby scaling with a convective overturn time that is (exactly)
proportional to the square-root of the stellar luminosity. One can think of
many other families of models that can potentially lead to this result, and
this opens room for more general descriptions of the magnetic dynamo.

\begin{acknowledgements}
  We thank Robert H.\ Cameron for illuminating discussions about magnetic
  dynamos. We acknowledge support from the DFG Collaborative Research Center
  SFB~963. AR acknowledges research funding from DFG grant RE 1664/9-1, VMP
  acknowledges funding from the DFG through GrK-1351.
\end{acknowledgements}

\bibliographystyle{apj}
\bibliography{refs}

\end{document}